\documentclass[12pt]{mn2e}
\usepackage{graphicx}
\usepackage{epsf}
\usepackage{lscape}
\usepackage{longtable}
\usepackage{rotating}
\usepackage{color}
\usepackage{amsfonts}

\newcommand{\apj}{ApJ}
\newcommand{\apjl}{ApJ}
\newcommand{\apjs}{ApJS}
\newcommand{\mnras}{MNRAS}

\begin{document}
\topmargin -0.5in 

\title[The Bright-end of the UV LF at $z\approx 7-9$ from CANDELS]{Constraining the Bright-end of the UV Luminosity Function for $z\approx 7-9$ Galaxies: results from CANDELS/GOODS-South}

\author[Silvio Lorenzoni, et al.\ ]  
{
Silvio Lorenzoni$^{1}$\thanks{E-mail: silvio.lorenzoni@astro.ox.ac.uk}, Andrew J. Bunker$^{1}$, Stephen M. Wilkins$^{1}$, 
\newauthor
 Joseph Caruana$^{1}$, Elizabeth R. Stanway$^{2}$, Matt J.\ Jarvis\,$^{3}$ \\
$^1$\,University of Oxford, Department of Physics, Denys Wilkinson Building, Keble Road, OX1 3RH, U.K. \\
$^{2}$\,Department of Physics, University of Warwick, Coventry, CV4 7AL, U.K.\\
$^{3}$\,Centre for Astrophysics, Science \& Technology Research Institute, University of Hertfordshire, Hatfield, Herts AL10\,9AB, U.K.}
\maketitle

\begin{abstract}
The recent {\em Hubble} Space Telescope near-infrared imaging with the Wide-Field Camera \#3 (WFC\,3) of the GOODS-South field in the CANDELS program covering nearly $100\,$arcmin$^2$, along with already existing Advanced Camera for Surveys optical data, makes possible the search for bright galaxy candidates at redshift $z\approx 7-9$ using the Lyman-break technique. We present the first analysis of $z'$-drop $z\approx 7$ candidate galaxies in this area, finding 19 objects. We also analyse $Y$-drops at $z\approx 8$, trebling the number of bright ($H_{AB} < 27$\,mag) $Y$-drops from our previous work, and compare our results with those of other groups based on the same data. The bright high redshift galaxy candidates we find serve to better constrain the bright end of the luminosity function at those redshift, and may also be more amenable to spectroscopic confirmation than the fainter ones presented in various previous work on the smaller fields (the {\em Hubble} Ultra Deep Field and the WFC\,3 Early Release Science observations).  We also look at the agreement with previous luminosity functions derived from WFC\,3 drop-out counts, finding a generally good agreement, except for the luminosity function of Yan et al.\ (2010) at $z\approx 8$, which is strongly ruled out.
\end{abstract} 


\section{Introduction}

Thanks to the installation of Wide Field Camera 3 (WFC3) on the {\em Hubble Space Telescope} ({\em HST}) in Summer 2009, the search for star-forming galaxies at redshifts $z \geq 7$ with the Lyman break technique (see Section~3) has become possible with the infrared channel of the Wide-Field Camera \#3 (WFC\,3) and led to the discovery of several galaxy candidates at $z\approx 7 - 10$.
From these candidates we can determine the rest frame UV luminosity function (LF) at these redshifts (Bunker et al.\ 2010, Wilkins et al.\ 2011a, Lorenzoni et al.\ 2011, Bouwens et al.\ 2011), an important tool in understanding the star formation history of the Universe, and also crucial to addressing the role of star-forming galaxies in reionization. These works show a broad agreement on the clear LF evolution from $z = 6$ (and below) to $z = 7$ with the characteristic luminosity $L^*$ fainter at higher redshifts, and suggest further evolution at even higher redshifts ($z\approx 8-10$), although based on fewer candidates.
The wealth of WFC\,3 data on the Great Observatories Origins Deep Survey South (GOODS-S) area recently obtained by The Cosmic Assembly Near-infrared Deep Extragalactic Legacy Survey (CANDELS, Grogin et al.\ 2011, Koekemoer et al.\ 2011), covering an area twice as large as the area surveyed in our previous papers, allows us to put better constraints on the bright end of the UVLF at $z\approx 7-9$. The larger field now available also allows the identification of brighter sources, which may be more amenable to spectroscopic follow-up. The fact that these new WFC\,3 images coincide with existing deep Adavanced Camera for Surveys (ACS) optical images is critical in rejecting potential interlopers -- the ACS filters lie below the Lyman limit and hence any detection at short wavelength will reject low redshift contaminants. This is a luxury not afforded to recent pure-parallel surveys for high-redshift drop-outs like the Hubble Infrared Pure Parallel Imaging Extragalactic Survey (HIPPIES, Yan et al.\ 2011) and the Brightest of Reionizing Galaxies (BoRG) survey (Trenti et al.\ 2011 \& Bradley et al.\ 2012).  In this paper we present for the first time a list of $z'$-drops at $z\approx 7$ drawn from the large CANDELS field of GOODS-S. We also present our selection of $z\approx 8$ $Y$-drops in this field, and compare this with recent independent analyses of CANDELS $Y$-drops in GOODS-S by Oesch et al.\ (2012) and Yan et al.\ (2012).

This paper is organised as follows: in Section~2 we outline the {\em HST} observations with WFC\,3 and the data reduction,
and in Section~3 we describe our colour selection to recover high-redshift Lyman break galaxies, and compare our
sample with those from other studies.  In Section~4 with discuss the UV luminosity function derived from the new data. Our conclusions are
presented in Section~5.
Throughout, we adopt the standard concordance cosmology of $\Omega_{M}= 0.3$, $\Omega_{\Lambda}= 0.7$ and use $H_0=70$\,km\,s$^{-1}$\,Mpc$^{-1}$. All magnitudes are on the AB system (Oke \& Gunn 1983).

\section{Observations and Data Reduction}

\subsection{Observations}

In this paper we analyse images from WFC\,3 on {\em HST} taken in the F105W, F125W and F160W filters, corresponding approximately to the near-infrared $Y$-, $J$- and $H$-bands.
The data come from the {\em HST} programs GO-12060, GO-12061 and GO-12062 in the CANDELS program (P.I.\ S.~Faber, {see Grogin et al.\ 2011; Koekemoer et al.\ 2011}), covering the areas of the GOODS-S field (Giavalisco et al.\ 2004) not covered by the Early Release Science (ERS) program GO/DD-11359 (P.I.\ R.~O'Connell, see Wilkins et al.\ 2010). The area is divided into a `deep' field, measuring $\sim 63\,$arcmin$^2$ with 3 orbits in each $Y_{105\mathrm{w}}$, $J_{125\mathrm{w}}$ and $H_{160\mathrm{w}}$ filters, and a `wide' field with one orbit per filter over an area of $\sim 33\,$arcmin$^2$ {(the areas quoted refer to the deepest area where the coverage has the maximum number of overlapping frames)}. Extensive ACS imaging has been carried in these areas in previous years (Giavalisco et al.\ 2004, Beckwith et al.\ 2006) in the $b$ (F425W), $v$ (F606W), $i$ (F814W) and $z'$ (F850LP) filters, allowing us to confidently use the Lyman-break technique to select likely high redshift star forming galaxies.

The infrared
channel of WFC\,3 was used, which is a Teledyne $1014\times 1014$ pixel HgCdTe detector (a 10-pixel strip on the edge is not illuminated by sky and used for pedestal estimation),
with a field of view of $123"\times 136"$.
The data were taken in ``MULTIACCUM" mode using SPARSAMPLE100, which non-destructively reads the array every 100\,seconds. These repeated non-destructive reads of the infrared array allow gradient-fitting to obtain the count rate (``sampling up the ramp'') and the flagging and rejection of cosmic ray strikes. 
In Table~\ref{tab:exptimes} we list the exposure time for both the `deep' and `wide' fields for each spectral band.

\begin{table*}
\begin{tabular}{lcccc}
& \multicolumn{3}{c}{WFC3 exposure times in ksec {(5$\sigma$ depth, AB mag)}} & Area\\
Field ID & $Y$-band & $J$-band &$H$-band & (arcmin$^2$) \\
\hline\hline
CANDELS DEEP & 8.1 ({27.8}) &7.4 ({27.3})  &7.7 {(27.2)} & 62.9\\
CANDELS WIDE  &2.7 {(26.8)} & 2.1 {(26.9)} &2.1 {(26.6)} & 32.8  \\
\end{tabular}
\caption{The total exposure time (in ksec) is listed for each WFC3 filter used in this study for both CANDELS `wide' and `deep' fields. {In parenthesis, the average depth for each filter over the area listed is shown. These are 5$\sigma$ limits calculated in apertures of $0\farcs6$ diameter, corrected as described in the text for aperture loss and reddening.}}
\label{tab:exptimes}
\end{table*}

\subsection{Data Reduction}

Data reduction is performed as described in our previous papers (Lorenzoni et al.\ 2011, Wilkins et al.\ 2011a).
We used the IRAF.STSDAS pipeline {\tt calwfc3}
to calculate the count rate and reject cosmic rays, then MULTIDRIZZLE
(Koekemoer et al.\ 2002) to combine exposures taking account of the geometric distortions and mapping on to an output
pixel size of $0\farcs06$ from an original $0\farcs13\,{\rm pix}^{-1}$, which corresponds to a $2\times 2$ block-averaging of the
GOODSv2.0 ACS drizzled images in $b$-, $v$-, $i$- and $z'$-bands. We used a MULTIDRIZZLE pixel fraction of 0.8 for the `deep' area and 1.0 for the `wide' area to recover some of the under-sampling. 
We used our own reduction of all the WFC3 data for the CANDELS GOODS-S `wide' area and of the $Y-$band data of the `deep' region. For the $J$- and $H$-bands covering the `deep', we used the reduced single epoch images made available by the CANDELS team\footnote{See {\tt http://candels.ucolick.org/data\_access/GOODS-S.html}}
and co-added these together with inverse-variance weighting (i.e.\ weighting each pixel by its exposure time).

For WFC3,
we use the zeropoints reported on {\tt http://www.stsci.edu/hst/wfc3/phot\_zp\_lbn}, last updated in January 2011,
where the zeropoints are 26.27, 26.25 \& 25.96 for F105W, F125W \& F160W. 

We perform photometry using fixed apertures of $0\farcs6$ diameter, {and introduce an aperture correction to account for the flux falling outside of the aperture. This correction was determined to be $\approx 0.2-0.25$\,mag in WFC3 from photometry with larger apertures on bright but unsaturated point sources. For the ACS images, the better resolution and finer pixel sampling require a smaller aperture correction of $\approx 0.1$\,mag. All the magnitudes reported in this paper have been corrected to approximate total magnitudes (valid for compact sources), and we have also corrected for the small amount of foreground Galactic extinction
toward these fields using the {\it COBE}/DIRBE \& {\it IRAS}/ISSA dust maps
of Schlegel, Finkbeiner \& Davis (1998). The optical reddening is
$E(B-V)=0.009$, equivalent to extinctions of
$A_{850lp}=0.012$,  $A_{105w}=0.010$, $A_{125w}=0.008$ \& $A_{160w}=0.005$.}


\subsection{Construction of Catalogues}

To perform the candidate selection we used the SExtractor photometry package (Bertin
\& Arnouts 1996), version 2.5.0. For $Y$-drops (objects clearly detected in the WFC3 $J$-band but with minimal flux in the $Y$-band and ACS images), apertures were `trained' in the $H$-band image, and running SExtractor in dual-image mode those apertures were used to measure the flux in the same locations in the $Y$-band and $J$-band images.
For each waveband we used a weight image derived from the exposure map.
The $z'$-drop selection was done from catalogs trained in the $J$-band rather than in the $H$-band

Tables~\ref{tab:zdropsd} and \ref{tab:ydropsd} present our photometry of $z'$- and $Y$-drops from SExtractor. The MULTIDRIZZLE geometric transformation and image re-gridding produces an output where the noise is highly correlated, hence measuring the standard deviation in blank areas of the final drizzled image will underestimate the noise (e.g., Casertano et al.\ 2000). 
As in our previous work we have corrected the magnitude errors returned by SExtractor using our ``true noise frames", combinations of the data obtained without using MULTIDRIZZLE and hence without correlation between adjacent pixels, to determine the scaling factor (typically SExtractor
underestimated the magnitude errors by a factor of $\approx2$ for pixfrac=0.8 used for most of our data). {We also measure the correlated noise (the standard deviation
of the background counts) in the drizzled image mosaics which we use for our source detection and photometry,
and use the relations in equation A13 of Casertano et al.\ (2000) to introduce a correction factor which depends on the output pixel scale 
and the size of the ``droplet" in the drizzling procedure (``pixfrac"). We generally found good agreement (at the 0.05\,mag level)
with our sensitivity measurements using the true-noise frames. The errors displayed in Tables \ref{tab:zdropsd} and Table \ref{tab:ydropsd} are the corrected output from SExtractor
}

\section{Candidate Selection}

Identification of candidates is achieved using the Lyman break
technique (e.g., Steidel et al. 1996), where a large colour
decrement is observed between filters either side of
 Lyman-$\alpha$  in the rest-frame of the galaxy. At $z>6$, the flux
 decrement comes principally from the large 
integrated optical depth of the intervening absorbers
(the Lyman-$\alpha$ forest).

\begin{figure}
\centering
\includegraphics[width=18pc]{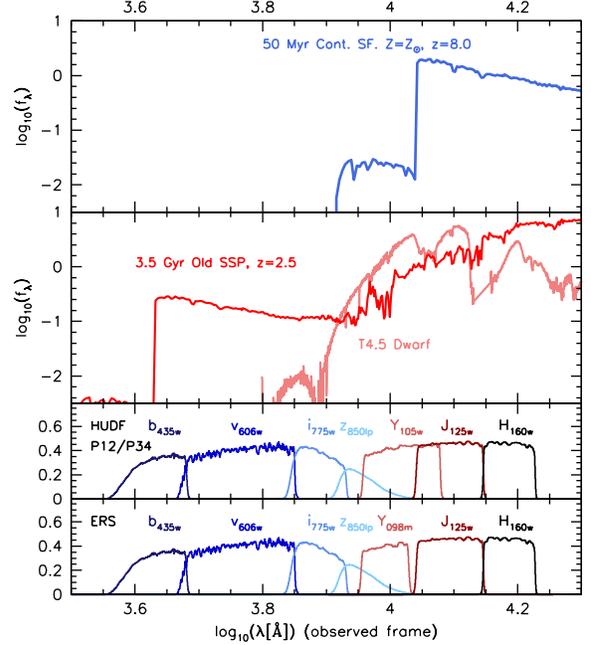}
\caption{{Top panel - Model (from the Starburst99, Leitherer et al. 1999) spectral energy distribution (SED) of a redshifted $z=8$ star forming galaxy. Middle panel - Potential contaminants: Observed SED of a low-mass dwarf star (class: $T4.5$, Knapp et al. 2004) together with the model (Starburst99) SED of a $3.5$Gyr Single-aged Stellar Population (SSP) at $z=2.5$. The bottom two panels show the transmission functions of the combination of filters available to each field.}}
\label{fig:lbg_spectra}
\end{figure}

At $z\approx 8-9$ the location of the Lyman-$\alpha$ break is redshifted to $\sim 1.1\mu$m -- the WFC3 $Y_{105\mathrm{w}}$ and $J_{125\mathrm{w}}$ are suitably located such that a $7.6<z<9.8$ star forming galaxy will experience a significant flux decrement between these two filters, while for $z\approx 7$ the break lies at $\sim 1\mu$m, between filters WFC3 $Y_{105\mathrm{w}}$ and ACS $z_{850\mathrm{lp}}$, with a redshift range of $6.5<z<8.0$ (see Figure \ref{fig:lbg_spectra}).
The selection efficiency drops at the extremes of these ranges. 

\subsection{Selection Criteria}

\begin{figure}
\centering
\includegraphics[width=17.7pc]{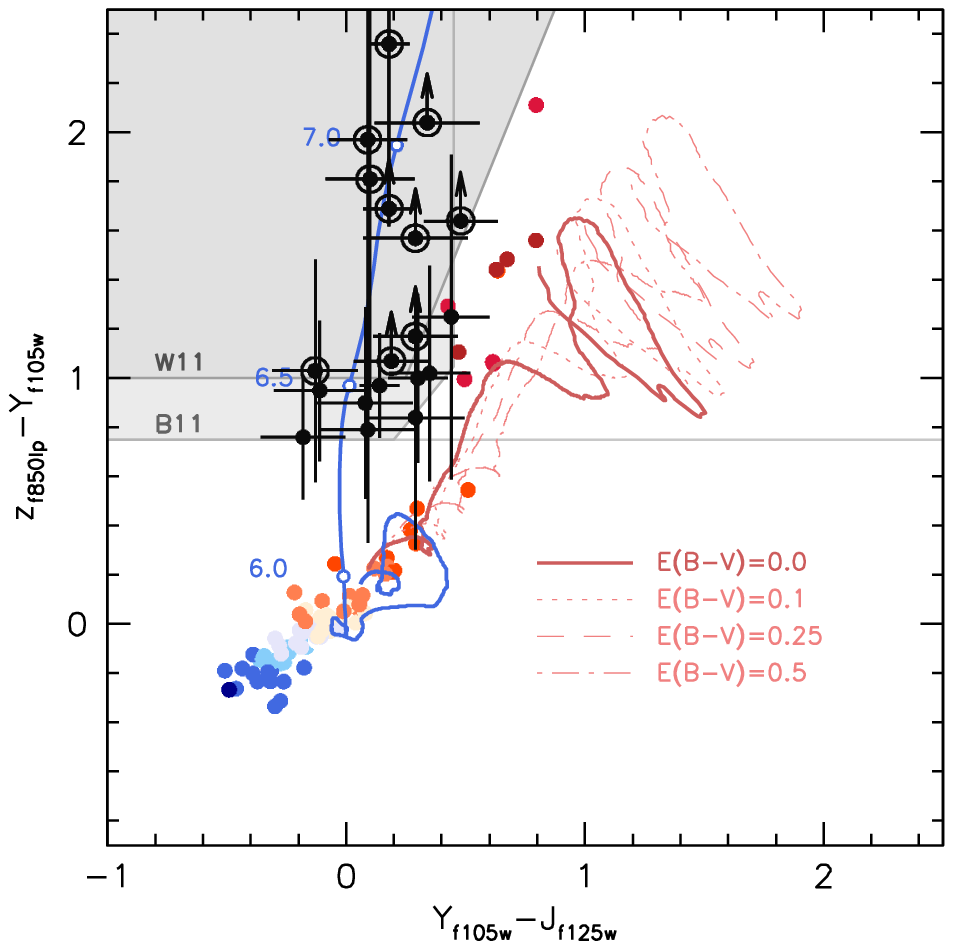}
\includegraphics[width=17.7pc]{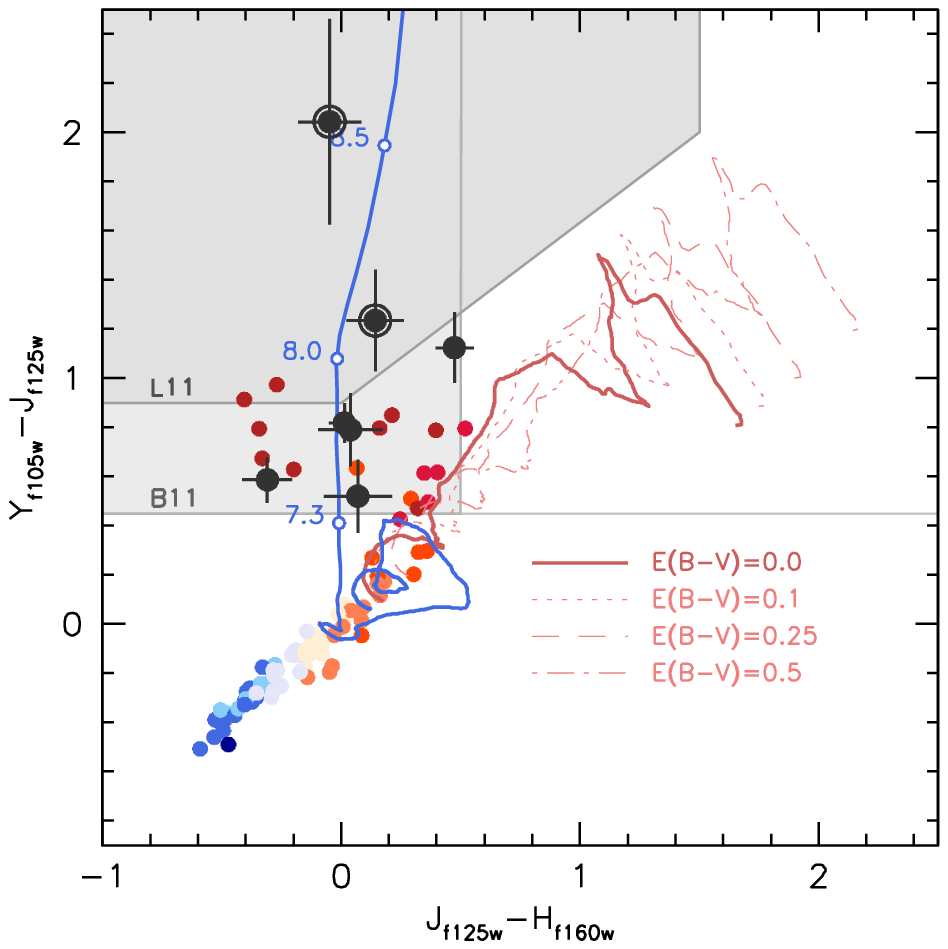}
\caption{Colour - colour diagrams for both $z'$-drops (top) and $Y$-drops (bottom). The shaded areas are the selection windows used, defined in Section 3.1 (light shading for the B11z and B11Y selection windows, darker shading for W11 and L11). The objects we found are shown as grey dots (objects not meeting any of the colour - colour windows we are considering), black dots (objects in B11z or B11Y) and black circled dots (objects meeting W11 or L11). The coloured dots denote the position of potential L and T dwarfs stars contaminants. The solid red line shows the colours that lower redshift galaxies (modelled as an instantaneous burst of star formation at $z=20$ and no dust) would have, {and the dotted, dashed and dot-dashed lines show this low-redshift template with reddenings of $E(B-V)=0.1$, $0.25$ \& $0.5$ respectively}. The blue line is the predicted path taken by high-redshift galaxies (constant star formation from $z=20$, no dust). For the red and blue tracks, numbers in correspondence with open circles indicate the redshift.}
\label{fig:cc_1}
\end{figure}

Our photometrically selected Lyman-break sample suffers from contamination due to photometric scatter and interlopers (in particular L and T type dwarf stars and red galaxies at intermediate redshift).
To discriminate candidates from these interlopers, we use the photometric data from another filter at wavelengths longer than the break, $J_{125\mathrm{w}}$ for $z'$-drops and $H_{160\mathrm{w}}$ for $Y$-drops and impose limitations on the $z_{850\mathrm{lp}}-Y_{105\mathrm{w}}$ and $J_{125\mathrm{w}}-H_{160\mathrm{w}}$ colours (respectively) as well, drawing a selection window in the colour-colour diagram that excludes most of the contaminants (Figure \ref{fig:cc_1}).

In this work we present objects within the colour -- colour windows we selected and with
detections of at least $5\,\sigma$ in the two bands at wavelengths longer than the Lyman-$\alpha$ break. 
Even though the selection windows rule out most of the intrinsically red interlopers, these can still be included in our selection because of photometric scatter.
To minimise this contamination, all objects with a $>2\sigma$ detection in any of the $b_{435\mathrm{w}}$, $v_{606\mathrm{w}}$ and $i_{775\mathrm{lp}}$ (below the Lyman limit) are classified as contaminants, ruling out in this way lower redshift red galaxies, which we expected to faintly detect in the optical bands (see Figure \ref{fig:lbg_spectra}).

Various colour selection windows have been proposed in the literature to remove contaminants and select high-redshift Lyman break galaxies. In this paper we use the criteria we derived previously for the $z'$-drops at $z\approx 7$ (Wilkins et al.\ 2011a, W11 hereafter):
{\begin{eqnarray*}
(z_{850lp}-Y_{105w})& > & 1.0\\
(z_{850lp}-Y_{105w})& > & 2.4\times (Y_{105w}-J_{125w})+0.9\\
(Y_{105w}-J_{125w}) & < & 1.0\\
\end{eqnarray*}}and for the $Y$-drops at $z\approx 8$ (Lorenzoni et al.\ 2011, L11 hereafter):
{\begin{eqnarray*}
(Y_{105w}-J_{125w})& > & 0.9\\
(Y_{105w}-J_{125w})& > & 0.73\times (J_{125w}-H_{160w})+0.9\\
(J_{125w}-H_{160w}) & < & 1.5\\
\end{eqnarray*}}
We also derive a list of candidates
obeying the colour-cuts proposed by Bouwens et al.\ (2011) for these redshifts (we label these Bouwens et al.\ criteria B11z for $z\approx 7$ and B11Y for $z\approx 8$ hereafter). This will allow for an easier comparison of candidates, and to investigate the effect of different selection windows on the derivation of a luminosity function.
For detections of less than $1\sigma$ in the $z'$- or $Y$-band, we quote a $1\sigma$ limit based on the noise and measured flux within the aperture.

\subsubsection{$z'$-drops}

\begin{figure}
\centering
\includegraphics[width=19.1pc]{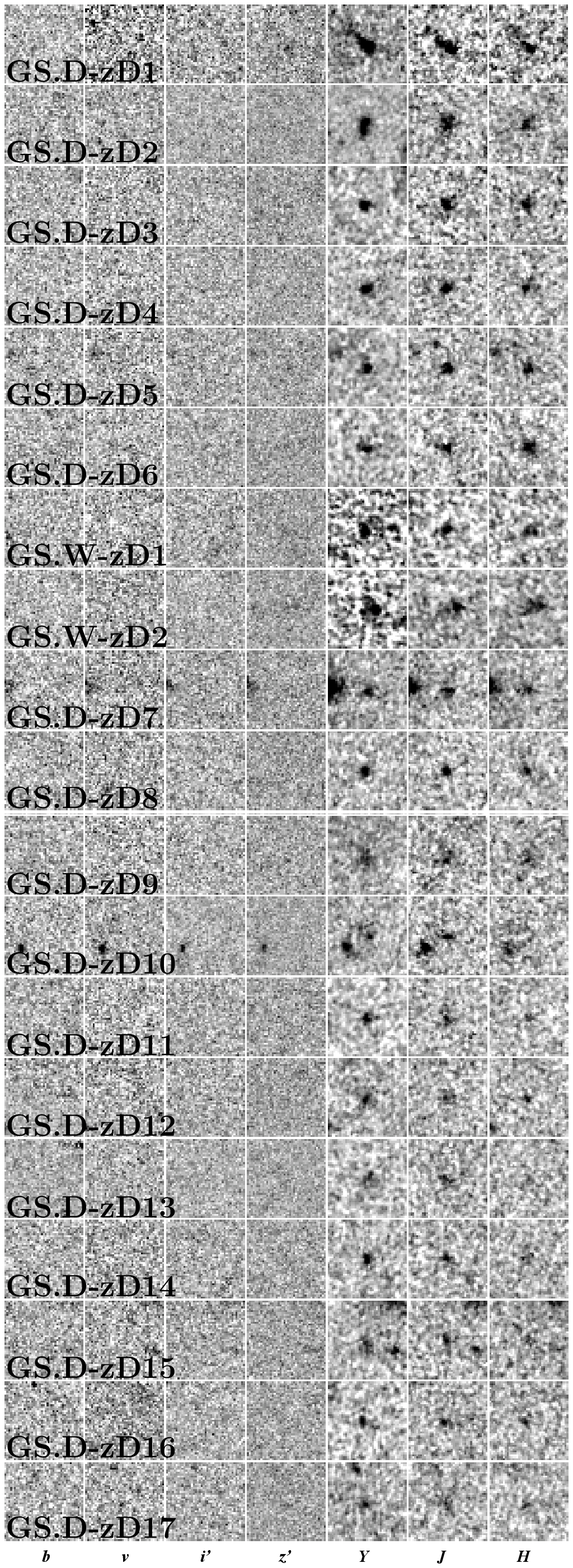}
\caption{$2\farcs4\times 2\farcs4$ $bvizYJH$ thumbnail images of potential $z\approx 7$ objects meeting our selection criteria in CANDELS GOODS-South field, ordered by $J$-band magnitude (brightest at the top).}
\label{fig:zstamps}
\end{figure}

In the `deep' area we find 17 objects meeting our selection criteria (see Figure \ref{fig:cc_1}, top panel). Of these, 16 candidates meet the B11z selection window, while 10 meet W11 (9 of which also match the B11z window). One of these objects is UDFz-4256656 from Bouwens et al.\ (2011) in the HUDF field. In the `wide' area, 2 objects meet the B11z window (GS.W-zD1 and GS.W-zD2), and none fall within the W11 colour selection.
Images of the $z'$-drops meeting the selection criteria (B11z and/or W11) are shown in Figure \ref{fig:zstamps}.
{Having three filters longwards of the break, it is possible to determine the UV spectral slope for the $z'$-drop candidates: the $Y$-band filter could be affected by either the Lyman break or Lyman-$\alpha$ emission, or both, so the $J$- and $H$-bands are necessary to have `clean' information on the UV slope. As in Wilkins et al.\ 2011b, $\beta$ is determined from the ($J_{125w}-H_{160w}$) colour by the relation $\beta = 4.28 \times (J_{125w}-H_{160w}) - 2.0$, which assumes that the slope is represented exactly by a power law. The $\beta$ values are listed in Table \ref{tab:zdropsd}: as already observed by Bunker et al.\ (2010), Wilkins et al.\ (2011b), Bouwens et al.\ (2010), the UV slopes of high-redshift galaxy candidates are very blue ($\beta \sim -2$), with fainter objects being bluer than the brighter. Note that the errorbars for faint candidates, due to photometric scatter, are considerable. }

\subsubsection{$Y$-drops}
\label{subsec:CandidateComparison}

\begin{figure}
\centering
\includegraphics[width=20pc]{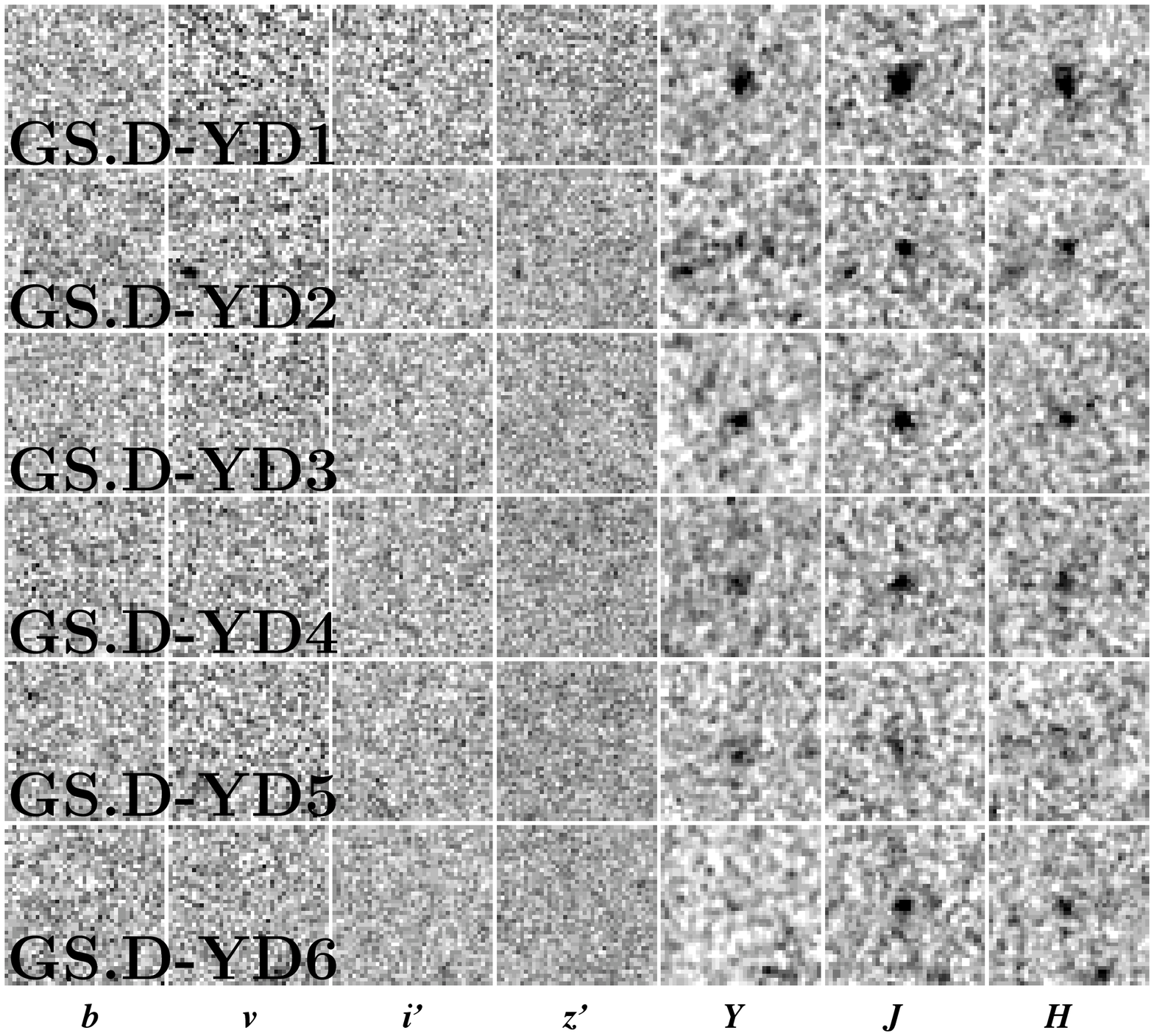}
\caption{$2\farcs4\times 2\farcs4$ $bvizYJH$ thumbnail images of potential $z\approx 8$ objects meeting our selection criteria in CANDELS GOODS-South field, ordered by $H$-band magnitude (brightest at the top).}
\label{fig:Ystamps}
\end{figure}

In the CANDELS `deep' area, Table \ref{tab:ydropsd}, 2 objects meet the L11 colour selection (Figure \ref{fig:Ystamps}), both of which are included in the 6 objects selected with the B11Y criteria (Figure \ref{fig:cc_1}, bottom panel). We did not find any $Y$-drop candidate in CANDELS `wide' area. 

\subsubsection{Comparison to Other Studies}

Both the `deep' and `wide' CANDELS observations of GOODS-S have been recently searched for $Y$-drop candidates by both Oesch et al.\ (2012, hereafter O12) and Yan et al.\ (2012, hereafter Y12), resulting in 11 and 8 high redshift galaxy candidates respectively.\\
Of the 11 O12 sources we match only 3 with our 6 candidates. Another object in our sample (GS.D-YD3) is also flagged as a {\em potential} candidate by O12 (CAND-2253348542) though is dismissed by O12 on the grounds of its stellar-like profile. 
We also match an additional 2 of our candidates with the 8 Y12 sources (there are no matches in common between all three candidate lists), 
thus all our candidates exist in either O12 or Y12.\\
Given the lack of agreement between the previous catalogs of $Y$-drops (O12 and Y12) with our new selection, and also the poor agreement between O12 and Y12 (there are 2 objects in common of which neither is in our candidate list) it is useful to examine each of the O12 and Y12 candidates in turn to identify why they were not selected by us.\\
Of the 8 O12 sources not selected as candidates by us, 2 objects, CANDY-2499448181 (which is also 048 in Y12) and CANDY-2209651371, are detected at $>2\sigma$ in a single optical band (though at $<3\sigma$). One object (CANDY-2320345371) is excluded because its $(Y-J)$ colour is slightly bluer than our selection window, while a further 4 sources fail to meet our S/N$>5$ criteria though do appear to be real objects (all detected at $>4\sigma$ in both $J$- and $H$-band). A single source (CANDY-219147298) is not matched within $0\farcs5$ of an object in our catalogue. Of the 6 Y12 sources not matched to our candidates the two brightest (048 and 100) are excluded on the basis of weak (2 to $3\sigma$) optical detections in a single band. The 4 remaining objects are excluded on the basis of S/N concerns (in that they fall below S/N$=5$ in one or both bands); in three cases (094, 035, 043) we detect the source at $>4\sigma$ in both $J_{125\mathrm{w}}$ and $H_{160\mathrm{w}}$ while the final object (085) is only detected at $2-3\sigma$ and has colours inconsistent with our selection window. \\ 
There are then two principal reasons for the Y12 and O12 objects being excluded from our candidate list; at the bright-end 2 objects in each study (with 1 in common) are excluded due to weak (2 to $3\sigma$) optical detections in single band; while at the faint end several sources are excluded on the basis of our S/N criteria. 
In all but one case (Y12: 085) these objects are detected at $>4\sigma$ in both $J_{125\mathrm{w}}$ and $H_{160\mathrm{w}}$ and have observed colours consistent with our selection window.  It then seems possible that some of the additional Y12 and O12 candidates are potential high-redshift star forming galaxies. However, these objects are nevertheless excluded from the subsequent analysis of the rest-frame UV luminosity function, as we want a robust sample. The computation of the effective volume takes into account our more conservative selection criteria, which should lead to the accurate luminosity function being recovered.

\begin{table*}
\begin{tabular}{lccccccccc}
\multicolumn{10}{c}{${\bf z\approx7}$} \\
\\
ID & RA & Dec & $z_{\rm AB}$ & $Y_{\rm AB}$ & $J_{\rm AB}$ & $H_{\rm AB}$ & $(z-$ & $(Y-$ & $\beta$ \\ 
& (J2000) & (J2000) & & & & & $Y)_{AB}$ & $J)_{AB}$ & \\
\hline\hline
GS.D-zD1 & 03:32:55.930 & -27:49:38.59 & $26.98 \pm 0.21$ & $26.01 \pm 0.044$ & $25.87 \pm 0.071$ & $26.25 \pm 0.15$ & $0.97$ & $0.14$ &  $-3.63 \pm 0.74$   \\ 
GS.D-zD2$^1$ & 03:32:37.181 & -27:48:56.68 & $28.88 \pm 0.74 $ & $26.52 \pm 0.043$ & $26.34 \pm 0.074$ & $26.46 \pm 0.12$ & $2.36$ & $0.18$ &  $-2.51 \pm 0.64$   \\ 
GS.D-zD3$^{1,2}$ & 03:32:08.130 & -27:46:40.88 & $> 28.49 $ & $26.85 \pm 0.12$ & $26.37 \pm 0.10$ & $26.39 \pm 0.15$ & $> 1.64$ & $0.48$ &  $-2.09 \pm 0.82$   \\ 
GS.D-zD4$^1$ & 03:32:36.006 & -27:44:41.74 & $> 28.26 $ & $26.57 \pm 0.067$ & $26.39 \pm 0.085$ & $26.52 \pm 0.14$ & $> 1.69$ & $0.18$ &  $-2.56 \pm 0.74$  \\ 
GS.D-zD5 & 03:32:25.447 & -27:50:53.36 & $27.76 \pm 0.33 $ & $26.76 \pm 0.094$ & $26.46 \pm 0.084$ & $26.44 \pm 0.12$ & $1.0$ & $0.3$ &   $-1.91 \pm 0.67$   \\ 
GS.D-zD6 & 03:32:09.583 & -27:46:32.06 & $28.01 \pm 0.42 $ & $26.99 \pm 0.13$ & $26.64 \pm 0.11$ & $26.01 \pm 0.10$ & $1.02$ & $0.35$ &  $0.70 \pm 0.73^*$   \\ 
GS.W-zD1 & 03:32:57.390 & -27:53:21.77 & $27.51 \pm 0.25 $ & $26.56 \pm 0.18$ & $26.67 \pm 0.16$ & $26.84 \pm 0.24$ & $0.95$ & $-0.11$ &  $-2.7 \pm 1.2$  \\ 
GS.W-zD2 & 03:32:36.729 & -27:54:42.12 & $27.35 \pm 0.21 $ & $26.59 \pm 0.17$ & $26.77 \pm 0.14$ & $26.58 \pm 0.17$ & $0.76$ & $-0.18$ &   $-1.19 \pm 0.94$  \\ 
GS.D-zD7 & 03:32:36.240 & -27:46:31.37 & $28.51 \pm 0.65 $ & $27.26 \pm 0.12$ & $26.82 \pm 0.11$ & $26.78 \pm 0.15$ & $1.25$ & $0.44$ &   $-1.83 \pm 0.85$  \\ 
GS.D-zD8$^1$ & 03:32:40.693 & -27:44:16.72 & $> 28.09 $ & $27.02 \pm 0.11$ & $26.83 \pm 0.12$ & $26.71 \pm 0.16$ & $> 1.07$ & $0.19$ &  $-1.49 \pm 0.91$  \\ 
GS.D-zD9$^1$ & 03:32:28.859 & -27:49:12.63 & $> 28.35 $ & $27.18 \pm 0.12$ & $26.89 \pm 0.14$ & $26.88 \pm 0.20$ & $> 1.17$ & $0.29$ &   $-1.96 \pm 1.10$ \\ 
GS.D-zD10$^1$ & 03:32:27.916 & -27:45:42.72 & $> 29.28 $ & $27.24 \pm 0.17$ & $26.9 \pm 0.14$ & $27.80 \pm 0.47$ & $> 2.04$ & $0.34$ &  $-5.8 \pm 2.1^*$  \\ 
GS.D-zD11$^1$ & 03:32:19.938 & -27:47:10.57 & $29.01 \pm 1.05 $ & $27.04 \pm 0.10$ & $26.95 \pm 0.13$ & $27.59 \pm 0.35$ & $1.97$ & $0.09$ & $-4.7 \pm 1.6$  \\ 
GS.D-zD12$^1$ & 03:32:47.638 & -27:48:29.21 & $28.98 \pm 0.93 $ & $27.17 \pm 0.11$ & $27.07 \pm 0.15$ & $27.66 \pm 0.39$ & $1.81$ & $0.1$ &  $-4.5 \pm 1.8$  \\ 
GS.D-zD13 & 03:32:12.512 & -27:47:56.86 & $28.12 \pm 0.37 $ & $27.22 \pm 0.12$ & $27.14 \pm 0.16$ & $27.87 \pm 0.46$ & $0.9$ & $0.08$ &  $-5.1 \pm 2.1$ \\ 
GS.D-zD14$^1$ & 03:32:37.230 & -27:45:38.41 & $28.05 \pm 0.44 $ & $27.02 \pm 0.10$ & $27.15 \pm 0.15$ & $27.36 \pm 0.26$ & $1.03$ & $-0.13$ & $-2.9 \pm 1.3$  \\ 
GS.D-zD15$^1$ & 03:32:30.793 & -27:50:27.19 & $> 29.03 $ & $27.46 \pm 0.17$ & $27.17 \pm 0.15$ & $27.56 \pm 0.32$ & $> 1.57$ & $0.29$ & $-3.7 \pm 1.6$  \\ 
GS.D-zD16 & 03:32:16.057 & -27:47:57.72 & $28.09 \pm 0.44$ & $27.3 \pm 0.13$ & $27.21 \pm 0.16$ & $27.70 \pm 0.37$ & $0.79$ & $0.09$ & $-4.1 \pm 1.8$  \\ 
GS.D-zD17 & 03:32:35.067 & -27:46:34.96 & $28.35 \pm 0.52 $ & $27.51 \pm 0.15$ & $27.22 \pm 0.15$ & $27.89 \pm 0.44$ & $0.84$ & $0.29$ &  $-4.8 \pm 2.0^*$ \\ 
\hline
\end{tabular}
$^{1}$ in W11 selection.; $^{2}$ not selected using B11 criteria.;
$^*$ outside the colour--colour selection window employed by Wilkins et al.\ (2011b) for a clean selection of $z$-drops for analysis of spectral slope, $\beta$.\\
\caption{$z'$-band drop out candidate at $z\approx7$ meeting either of the selection criteria described. Objects are ordered by apparent $J_{\rm AB}$ magnitude. Where quoted, limits are $1\sigma$}
\label{tab:zdropsd}
\end{table*}

\begin{table*}
\begin{tabular}{lcccccccll}
\multicolumn{10}{c}{${\bf z\approx8}$} \\
 \\
ID & RA & Dec & $Y_{\rm AB}$ & $J_{\rm AB}$ & $H_{\rm AB}$ & $(Y-J)_{AB}$ & $(J-H)_{AB}$ & $B11$ & $L11$ \\ 
\hline\hline
GS.D-YD1 & 03:32:48.921 & -27:47:07.36 & $27.0 \pm 0.11$ & $26.18 \pm 0.063$ & $26.17 \pm 0.077$ & $0.82$ & $0.01$ &  \checkmark &     \\ 
GS.D-YD2 & 03:32:14.135 & -27:48:28.96 & $28.18 \pm 0.3$ & $26.94 \pm 0.12$ & $26.8 \pm 0.13$ & $1.24$ & $0.14$ &  \checkmark & \checkmark \\ 
GS.D-YD3 & 03:32:25.330 & -27:48:54.07 & $27.18 \pm 0.11$ & $26.59 \pm 0.086$ & $26.9 \pm 0.13$ & $0.59$ & $-0.31$ &  \checkmark &     \\ 
GS.D-YD4 & 03:32:44.018 & -27:47:27.23 & $27.8 \pm 0.19$ & $27.01 \pm 0.13$ & $26.97 \pm 0.16$ & $0.79$ & $0.04$ &  \checkmark &     \\ 
GS.D-YD5 & 03:32:40.257 & -27:44:09.84 & $27.61 \pm 0.18$ & $27.09 \pm 0.14$ & $27.02 \pm 0.16$ & $0.52$ & $0.07$ &  \checkmark &     \\ 
GS.D-YD6 & 03:32:20.979 & -27:48:53.46 & $29.04 \pm 0.64$ & $27.0 \pm 0.13$ & $27.05 \pm 0.16$ & $2.04$ & $-0.05$ &  \checkmark & \checkmark \\ 
\hline
\end{tabular}
\caption{$Y$-band drop out candidate at $z\approx8$ meeting either of the selection criteria described. Objects are ordered by apparent $H_{\rm AB}$ magnitude.}
\label{tab:ydropsd}
\end{table*}

\begin{table}
\begin{tabular}{cccc}
Lorenzoni '12 & Oesh '12 & Yan '12 & Class \\
\hline\hline
 GS.D-YD1 & - & 064 & \\
 GS.D-YD2 & CANDY-2141348289 & - & \\
 GS.D-YD3 & CANDY-2253348542 & - & \\
 GS.D-YD4 & CANDY-2440247273 & - & \\
 GS.D-YD5 & - & 107& \\
 GS.D-YD6 &  CANDY-2209848535 & - & \\
 - & CANDY-2499448181 & 048 & O \\
 - & CANDY-2320345371 & - & W \\
 - & CANDY-2209651371 & - & O \\
 - & CANDY-2350049216 & 035 & F \\
 - & CANDY-2192147298 & - & ? \\
 - & CANDY-2181852456 & - & F \\
 - & CANDY-2379552208 & - & F \\
 - & CANDY-2408551569 & - & F \\
 - & - & 100 & O \\
 - & - & 094 & F \\
 - & - & 043 & F \\
 - & - & 085 & F, W \\
\end{tabular}
 \\
 \\
F - Object too faint in $J-$ and/or $H-$band for our selection criteria.\\
O - Detection of more than 2\,$\sigma$ in at least one of the optical bands.\\
W - Object outside our colour-colour selection windows.\\
 ? \, - Object not picked up by SExtractor.\\
\caption{We list here candidates identified by Oesch et al.\ (2012) and Yan et al.\ (2012), second and third column respectively, and match them with ours when possible (first column) or give the reason why we do not find them (fourth column).}
\label{}
\end{table}

\section{Discussion}

\subsection{The bright-end of the UV luminosity function at $z\approx 7-9$ from CANDELS}
\label{sec:lumfunc}

From our selection of $z'$- and $Y$-drops we can recover the volume density of galaxies at $z\approx 7$ and $z\approx 8$ as a function of the rest-frame UV luminosity.
The Lyman-break technique does not have uniform sensitivity on the probed redshift range, so we quantify the probability of recovering a high-redshift galaxy in our survey as a function of redshift and absolute UV magnitudes, $p(M_{UV},z)$, {with simulations. To perform these simulations we add into the images a large number of fake galaxies, with properties similar to those of the observed high-redshift population
(i.e. compact with half-light radii $r_{hl}\approx 0\farcs1$, large Lyman-$\alpha$ forest decrement of $D_A\approx 0.99$ and blue rest-frame UV colours). We then run our selection procedure and infer the probability of recovering such galaxies as a function of redshift and magnitude.} 
{From this probability} the effective survey volume $V_{eff}$ can be calculated, with the same approach described in Steidel et al.\ (1999) and Stanway, Bunker \& McMahon (2003).
We assume the LF to have a Schechter\ (1976) profile with four fixed values for $\alpha$, -1.5, -1.7, -1.9 and -2.1, as the faint end slope cannot be strongly constrained with current data. The other Schechter parameters, $\phi^{*}$ and $M_{1600}^{*}$, are determined by maximising the Poissonian likelihood of observing a number of objects in a magnitude bin.

\begin{figure}
\centering
\includegraphics[width=20pc]{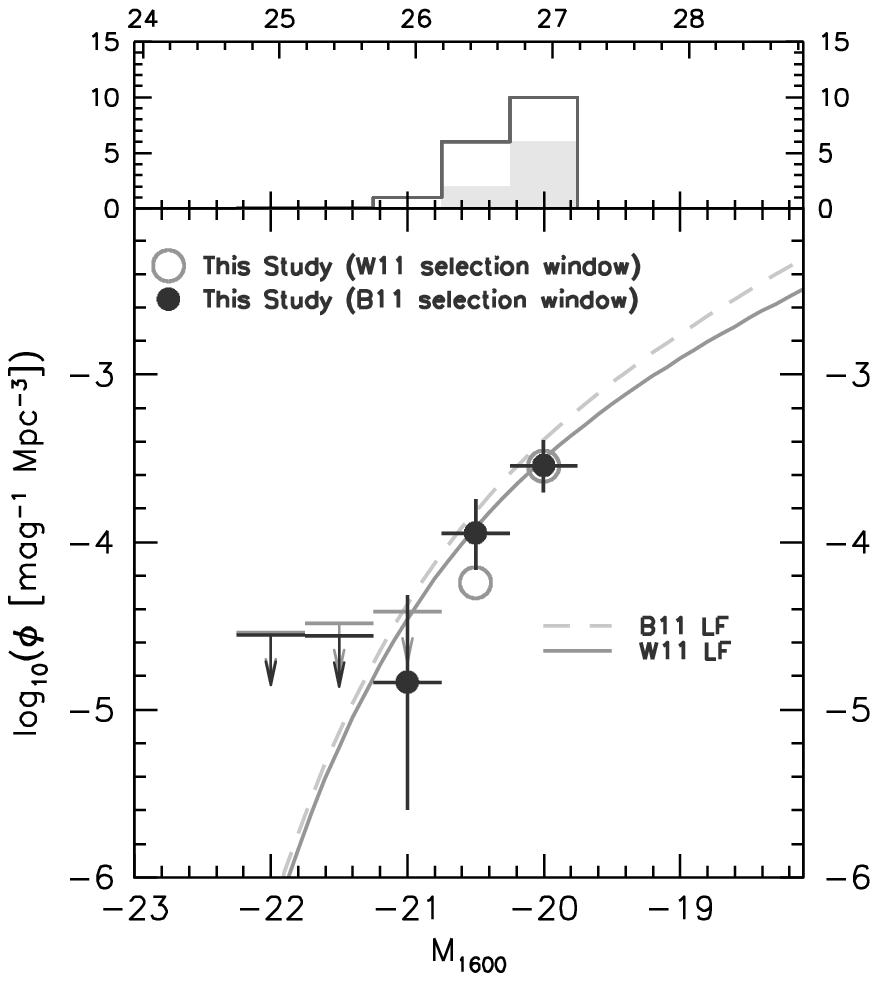}
\caption{The luminosity distribution (top) and luminosity function (bottom) of $z'$-drop selected sources at $z\approx 7$. 
{Our datapoints are plotted against Wilkins et al.\ (2011a, solid line) and Bouwens et al.\ (2011, dashed line) luminosity functions.}
The uncertainty bars represent the $68.2\%$ poissonian confidence interval of the number density $\phi$. The upper limits denote the maximum value of the $68.2\%$ confidence interval with $n=0$ observations. This corresponds roughly to $n=1.84$, i.e. the for an observed $n=0$ there is a $68.2\%$ chance the true value is $<1.84$.}
\label{fig:lf_z}
\end{figure}

\begin{figure}
\centering
\includegraphics[width=20pc]{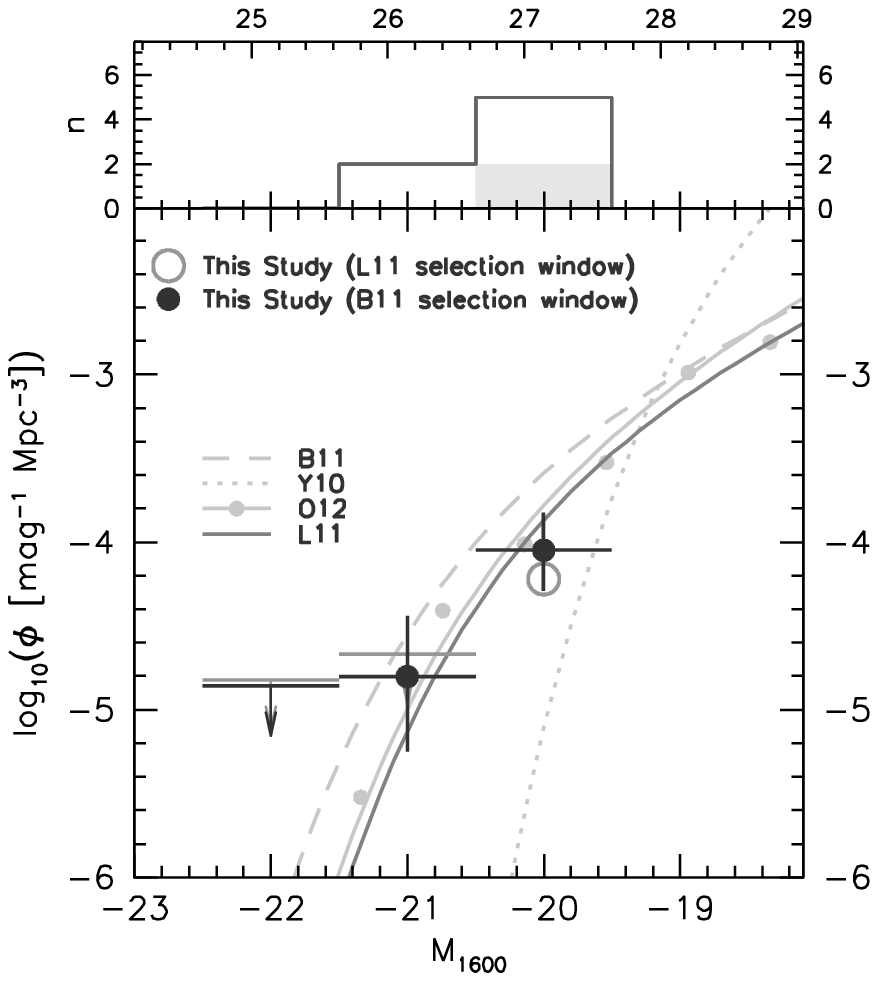}
\caption{The luminosity distribution (top) and luminosity function (bottom) of $Y$-drop selected sources at $z\approx 8$. 
{Our datapoints are plotted against several luminosity functions: Lorenzoni et al.\ (2011, solid dark line), Oesch et al.\ (2012, solid light line), Bouwens et al.\ (2011, dashed line) and Yan et al.\ (2010, dotted line).}
The uncertainty bars represent the $68.2\%$ poissonian confidence interval of the number density $\phi$. The upper limits denote the maximum value of the $68.2\%$ confidence interval with $n=0$ observations. This corresponds roughly to $n=1.84$, i.e. the for an observed $n=0$ there is a $68.2\%$ chance the true value is $<1.84$.}
\label{fig:lf_Y}
\end{figure}

In Figures~\ref{fig:lf_z} and \ref{fig:lf_Y} we plot our datapoints at $z\approx 7$ (W11 selection window) and $z\approx 8$ (L11 selection window), respectively, against several luminosity functions from our previous work (Wilkins et al.\ 2011a, Lorenzoni et al.\ 2011) and other publications (Bouwens et al.\ 2011, Oesch et al.\ 2012, Yan et al.\ 2010).
In the same Figures we also plot our datapoints obtained for the B11z and B11Y selection windows. As can be clearly seen, the number densities inferred from the different selection windows are in good agreement, within the error bars.
{We will therefore consider the lists of candidates obtained using the B11z and B11Y selection windows.}

\begin{figure}
\centering
\includegraphics[width=20pc]{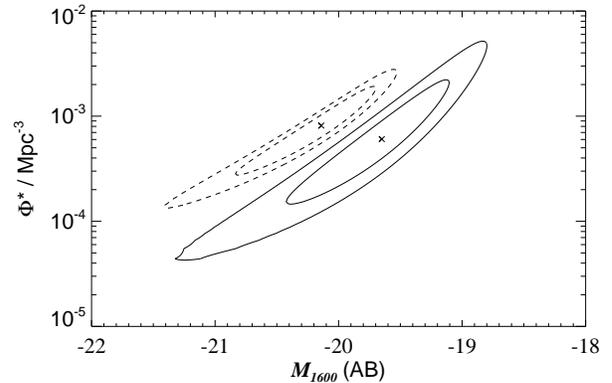}
\caption{{The likelihood contours for the luminosity function of $z'$-drops (dashed lines) and $Y$-drops (solid lines), showing the correlation between the fitted $M^*$ and $\phi^*$ parameters for a Schechter function fit, using our sample of galaxies from the B11 colour selection. A faint-end slope of $\alpha=-1.9$ is adopted here. The 68\% (inner) and 95\% (outer) likelihood contours are shown.The cross represents the best-fit parameter values.}}
\label{fig:LFcontours}
\end{figure}

In Table~\ref{tab:7fit} and Table~\ref{tab:8fit} we show the best fitting results for the LF at redshifts $z\approx 7$ and $z\approx 8$, respectively, for each of the selections windows used. The candidates found in our previous works in the HUDF and ERS fields (Wilkins et al.\ 2011a, Lorenzoni et al.\ 2011) are also included in all the LF calculations. {In fitting the Schechter luminosity function, $phi^*$ and $M^*$ are highly correlated, so we show the error ellipses ($1\,\sigma$ and $2\,\sigma$ significance contours) for the $z\approx 7$ and $z\approx 8-9$ luminosity functions in Figure~\ref{fig:LFcontours}.}

We note very good agreement at $z\approx 7$ between the best fitting LFs obtained using the two different selection windows (W11 and B11z). These results are also in line with our previous estimates (Wilkins et al.\ 2011a).

At $z\approx 8$, the L11 selection window adds only 2 candidates to our previous sample of $Y$-drops.
The B11Y selection yields 6 candidates, and combining these with our previous sample (accounting
for the different effective volumes probed by the colour selections) produces luminosity functions
(Table~\ref{tab:8fit}) consistent with those of several previous studies {Lorenzoni et al.\ (2011), Bouwens et al.\ (2011), Oesch et al.\ (2012)}, which indicate
fainter characteristic luminosity, $L^*$, than at lower redshifts. However, these results at $z\approx 8$
are strongly inconsistent with the LF proposed by Yan et al.\ (2010) on the basis of their analysis of the HUDF, in which they claimed far more faint $Y$-drop galaxies than in the analyses of other groups (Bunker et al.\ 2010; McLure et al.\ 2010; Bouwens at el.\ 2010). As can be seen in Figure~\ref{fig:lf_Y}, our measured number densities of $Y$-drops at brighter magnitudes ($M_{UV}=-21$ \& $-20$) are inconsistent by an order of magnitude or more than the expectation from the Yan et al.\ (2010) LF. 


\begin{figure}
\centering
\includegraphics[width=22pc]{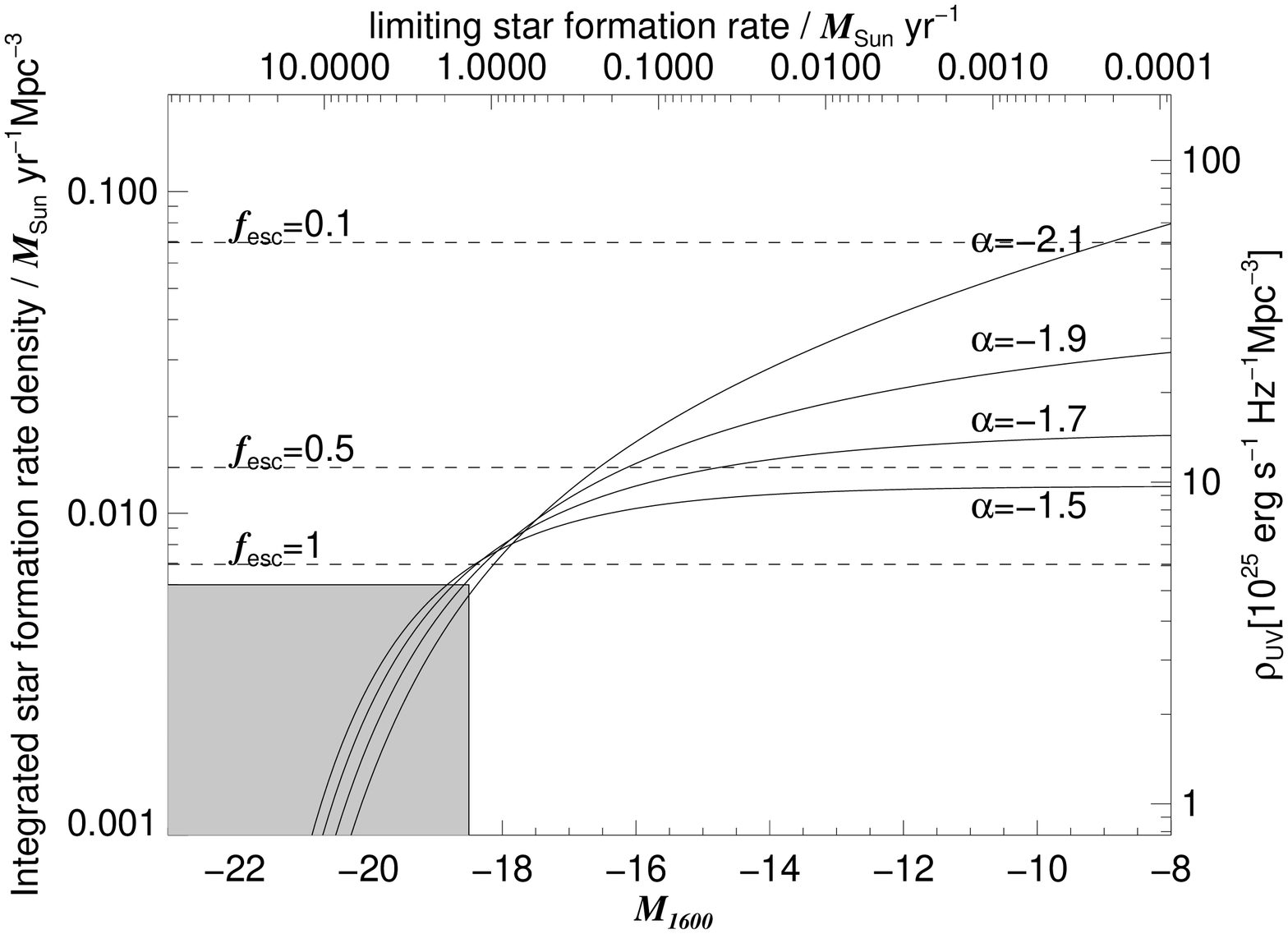}
\caption{The solid lines are the total star formation rate density (left axis) or ionising flux density (right axis)
inferred from the luminosity function fits for our $z'$-drop sample (for faint end slopes
$\alpha=[-1.5,-1.7,-1.9,-2.1]$), integrating down to the limiting absolute
magnitude in the rest-frame UV shown on the lower $x$-axis (in AB magnitudes); the upper $x$-axis
shows the equivalent unobscured star formation rate. The dashed lines show the requirement to keep
the Universe ionised at $z=7$, using the relation from Madau, Haardt \& Rees (1999) and assuming
a low clumping factor of $C=5$. We show the requirements for escape fractions of $f_{\rm esc} =0.1$, $0.5$ \& $1$.
Where the solid lines cross the dashed lines, reionzation can be achieved. The shaded region is where the
current deepest observations probe (the HUDF).}
\label{fig:reionization_zdrops}
\end{figure}

\begin{figure}
\centering
\includegraphics[width=22pc]{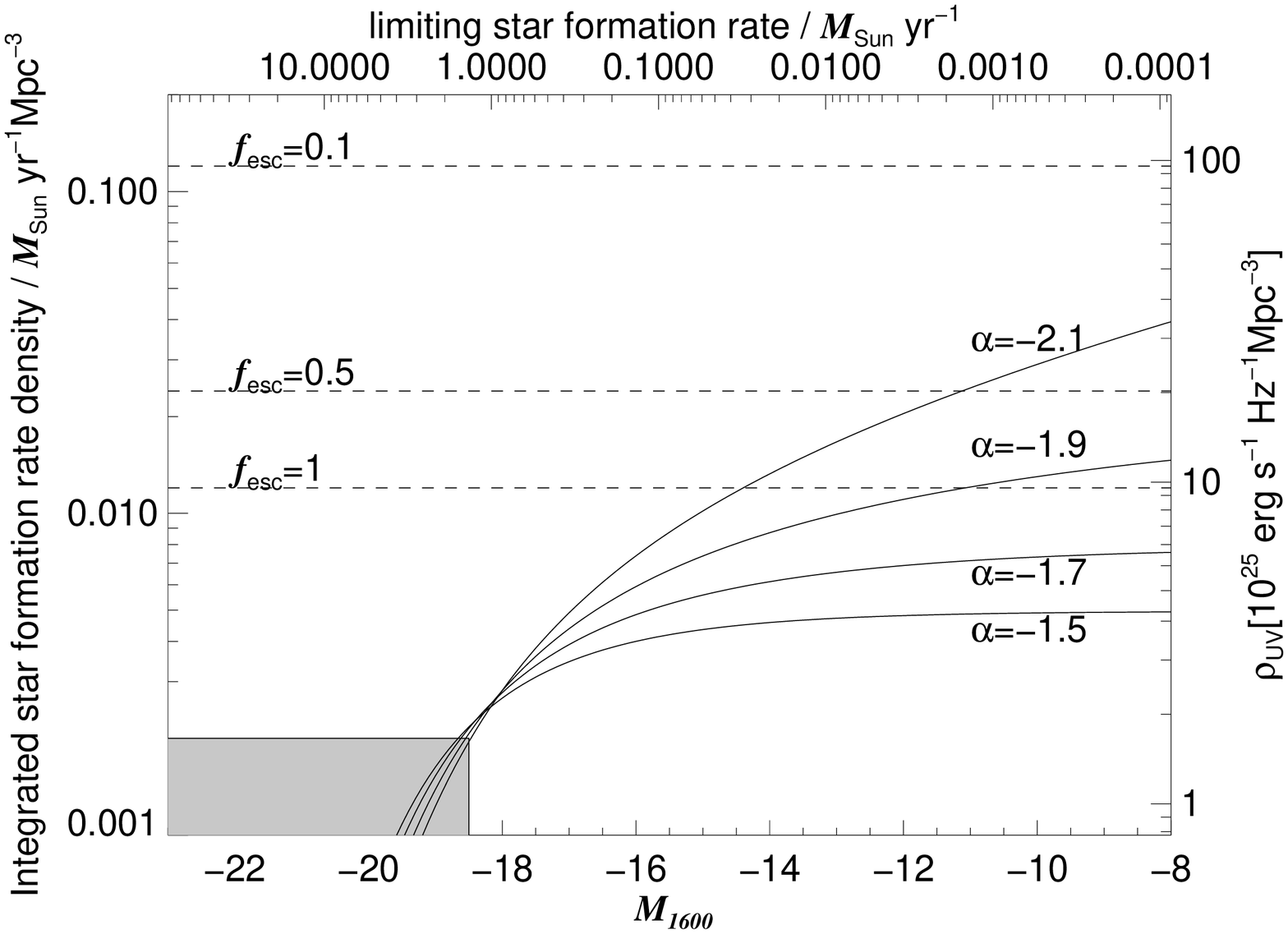}
\caption{The solid lines are the total star formation rate density (left axis) or ionising flux density (right axis)
inferred from the luminosity function fits for our $Y$-drop sample (for faint end slopes
$\alpha=[-1.5,-1.7,-1.9,-2.1]$), integrating down to the limiting absolute
magnitude in the rest-frame UV shown on the lower $x$-axis (in AB magnitudes); the upper $x$-axis
shows the equivalent unobscured star formation rate. The dashed lines show the requirement to keep
the Universe ionised at $z=8.6$, using the relation from Madau, Haardt \& Rees (1999) and assuming
a low clumping factor of $C=5$. We show the requirements for escape fractions of $f_{\rm esc} =0.1$, $0.5$ \& $1$.
Where the solid lines cross the dashed lines, reionzation can be achieved. The shaded region is where the
current deepest observations probe (the HUDF).}
\label{fig:reionization_Ydrops}
\end{figure}

{We now compare the star formation rate (SFR) densities obtained by integrating the $z\approx7$ and $z\approx8$ luminosity functions  down to various limiting magnitudes (Figures~\ref{fig:reionization_zdrops} \& \ref{fig:reionization_Ydrops} to the SFR densities required for reionization from the Madau, Haardt \& Rees (1999) relation:
\[
{\dot{\rho}}_{\mathrm{SFR}}\approx \frac{0.012\,M_{\odot}\,{\mathrm{yr}}^{-1}\,{\mathrm{Mpc}}^{
-3}}{f_{\mathrm{esc}}}\,\left( \frac{1+z}{1+8.6}\right) ^{3}\,\left( \frac{\Omega_{b}\,h^
2_{70}}{0.0462}\right) ^{2}\,\left( \frac{C}{5}\right)
\]
We have updated equation 27 of Madau, Haardt \& Rees (1999) for a more recent concordance cosmology estimate of the baryon
density from Larson et al.\ (2010), $\Omega_b\,h_{100}^2=0.022622$. 
In the above equation, $C$ is the clumping factor of neutral
hydrogen, $C=\left< \rho^{2}_{\mathrm{HI}}\right> \left< \rho_{\mathrm{HI}}\right> ^{-2}$, whose used value in this work is 5 (Pawlik et al.\ 2009). $f_{\mathrm{esc}}$ is the escape fraction of ionizing photons, which is highly uncertain - we consider escape fractions as high as 100 per cent (rather implausible)
and down to 10 per cent (which may be the average at $z\approx 3$ population, Nestor et al.\ 2011). At $z\approx 8.6$ (the average redshift of the $Y$-drops),
reionization cannot be achieved with the observed luminosity functions unless the slope is $\alpha=-1.9$ or steeper, even if the escape fraction is 100 per cent.
However, a steeper faint end slope, an even lower IGM clumping factor, and a low-metallicity population (or a top-heavy IMF) might still provide sufficient photons for star-forming galaxies to reionize the Universe (see Lorenzoni et al.\ 2011).
}

\begin{table*}
\begin{tabular}{ccccc}
\multicolumn{5}{c}{${\bf z\approx7}$} \\
 \\
\multicolumn{1}{c}{ } & \multicolumn{2}{c}{W11} & \multicolumn{2}{c}{B11z}\\
$\alpha$ & $M_{1600}^{*}$ [AB mag] & $\phi^{*}$ [Mpc$^{-3}$] & $M_{1600}^{*}$ [AB mag] & [Mpc$^{-3}$] \\
\hline\hline
$-$1.5 & $-$19.75 & 0.00152 & $-$19.75 & 0.00159 \\
$-$1.7 & $-$19.95 & 0.00110 &  $-$19.93 & 0.00119 \\
$-$1.9 & $-$20.19& 0.00072 & $-$20.14 & 0.00081 \\
$-$2.1 & $-$20.51& 0.00039 & $-$20.40 & 0.00049 \\
\end{tabular}
\caption{The best fit values for $M_{1600}^{*}$ and $\phi^{*}$ at $z\approx7$ for a Schechter function assuming fixed $\alpha\in\{-1.5,-1.7,-1.9,-2.1\}$ for both the W11 (columns 2 and 3) and B11z (columns 4 and 5) selection windows.}
\label{tab:7fit}
\end{table*}

\begin{table*}
\begin{tabular}{ccccc}
\multicolumn{5}{c}{${\bf z\approx8}$} \\
 \\
\multicolumn{1}{c}{ } & \multicolumn{2}{c}{L11} & \multicolumn{2}{c}{B11Y}\\
$\alpha$ & $M_{1600}^{*}$ [AB mag] & $\phi^{*}$ [Mpc$^{-3}$] & $M_{1600}^{*}$ [AB mag] & [Mpc$^{-3}$]  \\
\hline\hline
$-$1.5 & $-$19.10 & 0.00143 & $-$19.42 & 0.00088 \\
$-$1.7 & $-$19.23 & 0.00119 &  $-$19.53 & 0.00075 \\
$-$1.9 & $-$19.37& 0.00095 & $-$19.66 & 0.00060 \\
$-$2.1 & $-$19.54& 0.00069 & $-$19.80 & 0.00046 \\
\end{tabular}
\caption{The best fit values for $M_{1600}^{*}$ and $\phi^{*}$ at $z\approx8$ for a Schechter function assuming fixed $\alpha\in\{-1.5,-1.7,-1.9,-2.1\}$ for both the L11 (columns 2 and 3) and B11Y (columns 4 and 5) selection windows.}
\label{tab:8fit}
\end{table*}

\section{Conclusions}

In this paper we present a list of candidate high-redshift star-forming galaxies identified with the Lyman-break technique using {\em HST}/WFC3 near-infrared data within the CANDELS programme.
We have presented the first analysis of $z'$-drop candidate galaxies at $z\approx 7$ images with {\em HST}/WFC3 in the new CANDELS imaging of the GOODS-S field, building on previous
work by our team (Bunker et al.\ 2010; Wilkins et al.\ 2010, 2011a) in the smaller HUDF and ERS fields within GOODS-S. We also use the colour selections derived by Lorenzoni et al.\ (2011) and Bouwens et al.\ (2011) to identify candidate $z\approx 8$ $Y$-drops galaxies in this field, and compare our catalogues with those independently derived from the same CANDELS field by Oesch et al.\ (2012) and Yan et al.\ (2012).
We treble the number of bright ($H_{mag} < 27$) $Y$-drops from Lorenzoni et al.\ (2011) and double the number of bright ($J_{mag} < 27.2$) $z'$-drops from Wilkins et al.\ (2011a). 

The bright high redshift galaxy candidates we found serve to better constrain the bright end of the luminosity function at those redshift, and may also be more amenable to spectroscopic confirmation than the fainter ones presented in various previous work on the smaller fields (HUDF and ERS). Indeed, with AB magnitudes of $\approx 26$ (longward of the break),
we could hope to detect Lyman-$\alpha$ emission lines with rest
frame equivalent widths of a few tens of \AA ngstroms (typical
of Lyman break galaxies at $z\sim 3-6$, e.g. Stanway et al.\ 2004) in $\approx 5$\,hours'
spectroscopy with an instrument such as XSHOOTER on VLT (see Caruana
et al.\ 2012). If spectroscopy reveals that Lyman-$\alpha$
does not emerge at these redshifts, then our bright Lyman-break galaxy sample can potentially
place strong constraints on the absorption of the Gunn-Peterson (1965) damping wing (and hence the neutral
fraction of hydrogen at $z\sim 8$).

We also look at the agreement with previous luminosity functions derived from WFC3 drop-out counts, and find good agreement with those of Wilkins et al.\ (2011a) and Bouwens et al.\ (2011) at $z\approx 7$, and Lorenzoni et al.\ (2011) and Oesch et al.\ (2012) at $z\approx 8$. However, our results strongly rule out the $z\approx 8$ luminosity function proposed by Yan et al.\ (2010).

\subsection*{Acknowledgements}
Based on observations made with the NASA/ESA Hubble Space Telescope,
obtained from the Data Archive at the Space Telescope Science Institute, which is operated by the Association
of Universities for Research in Astronomy, Inc., under NASA contract
NAS 5-26555. These observations are associated with programmes \#GO-12060, \#GO-12061 and \#GO-12062. We are grateful to
the CANDELS team for making their data reductions public.
MJJ acknowledges the support of a RCUK
fellowship. 
SL and JC are supported by the Marie Curie Initial Training Network ELIXIR of the European Commission under
contract PITN-GA-2008-214227. AB and SW acknowledge support from a STFC standard grant ST/G001774/1.
We thank the anonymous referee for constructive comments on this paper.

\bsp

\end{document}